\documentclass[10pt]{article}

\usepackage{url, graphicx, color, varioref, amsfonts, longtable, float}

\title{Representation of Boolean Quantum Circuits as Reed-Muller Expansions}

\author{
Ahmed Younes \and Julian Miller\\
\and\\
School of Computer Science\\
The University of Birmingham\\ 
Birmingham\\
B15 2TT\\
United Kingdom\\
\{A.Younes , J.Miller\}@cs.bham.ac.uk \\
}

\begin{document}
\maketitle
\begin{abstract}
In this paper we show that there is a direct correspondence between quantum Boolean operations and 
certain forms of classical (non-quantum) logic known as Reed-Muller expansions. This allows us to 
readily convert Boolean circuits into their quantum equivalents. A direct result of this is that the 
problem of synthesis and optimization of quantum Boolean logic can be tackled within the field of 
Reed-Muller logic.
\end{abstract}

\section{Introduction}

Implementing Boolean functions on quantum computers is an essential aim, in 
the exploration of the benefits, which may be gained from systems operating 
by quantum rules. It is important to find the corresponding quantum 
circuits \cite{yao93}, which can carry out the operations we use to implement on conventional 
computers. On classical computers, a circuit can be built for 
any Boolean function using AND, OR and NOT gates. This set of gates cannot, 
in general be used to build quantum circuits because the operations are not 
reversible \cite{toffoli80}. A corresponding set of reversible gates must be used to 
build a quantum circuit for any Boolean operation. In classical computer 
science, many clever methods have been used to obtain more efficient digital 
circuits \cite{old:bok} for a given Boolean function. Recently, there have been efforts 
to find an automatic way to create efficient quantum circuits implementing 
Boolean functions. A method proposed in \cite{practmethod} used a modified version of 
\textit{Karnaugh maps} \cite{old:bok} and depends on a clever choice of certain minterms to be used in 
minimization process, however it appears that this method has poor 
scalability. Another work \cite{transrules}, includes a very useful set of transformations 
for quantum Boolean circuits and proposes a method for building quantum 
circuits for Boolean functions by using extra auxiliary qubits, however, 
this will increase the number of qubits to be used in the final circuits. In 
previous work \cite{younes1}, we showed a method by which we can convert a truth table 
of any given Boolean function to its quantum Boolean circuit by applying a 
set of transformations after which we get the final circuit. In this paper we will show that there is 
a close connection between quantum Boolean operations and certain classical Boolean operations known as 
Reed-Muller logic expansions \cite{bookRM}. This means that the study of synthesis and optimization of 
quantum Boolean logic can be carried out in the classical Reed-Muller logic domain.

The plan of the paper is as follows: In section 2, we review the principles of classical Reed-Muller logic. 
In section 3, we discuss the principles of quantum Boolean logic. In section 4, we show how we may 
implement quantum Boolean logic circuits directly from the corresponding classical Reed-Muller expansions. 
The paper ends with some conclusions and suggestions for further investigations.

\section{Reed-Muller Expansions (RM)}

In digital logic design two paradigms have been studied. The first uses the operations of AND, OR and NOT 
and called {\it canonical Boolean logic}. The second used the operations AND, XOR and NOT and called 
{\it Reed-Muller logic} (RM). RM is equivalent to modulo-2 algebra. In this section we review the 
properties of RM logic. 

\subsection{Modulo--2 Algebra}

For any Boolean variable $x$, we can write the following $XOR$ expressions:

\smallskip

$ {\begin{array}{*{20}c}
 {x \oplus 1 = \overline x ,} \hfill & {x \oplus 0 = x} \hfill \\
 {\overline x \oplus 1 = x,} \hfill & {\overline x \oplus 0 = \overline x } \hfill \\ 
\end{array} }$

\smallskip

Let $\mathop x\limits^\bullet $ be a variable representing a Boolean 
variable in its true $(x)$ or complemented form $(\overline x )$, then we can 
write the following expressions:

\smallskip

$\begin{array}{l}
 {\begin{array}{*{20}c}
 {\mathop x\limits^\bullet \oplus 1 = \overline {\mathop x\limits^\bullet } 
,} \hfill & {\mathop x\limits^\bullet \oplus 0 = \mathop x\limits^\bullet } 
\hfill \\
 {\mathop x\limits^\bullet \oplus \mathop x\limits^\bullet = 0,} \hfill & 
{\mathop x\limits^\bullet \oplus \overline {\mathop x\limits^\bullet } = 1} 
\hfill \\
\end{array} } \\ 
 {\begin{array}{*{20}c}
 {1 \oplus 1 = 0,} \hfill & {\mathop {x_0 }\limits^\bullet (1 \oplus \mathop 
{x_1 }\limits^\bullet ) = \mathop {x_0 }\limits^\bullet \oplus \mathop {x_0 
}\limits^\bullet \mathop {x_1 }\limits^\bullet } \hfill \\
\end{array} } \\ 
 f \oplus f\mathop x\limits^\bullet = f\overline {\mathop x\limits^\bullet } 
,\mbox{where }f\mbox{ is any Boolean function.} \\ 
 \end{array}$

\smallskip

For any $XOR$ expression, the following properties hold:

\smallskip

\begin{itemize}
\item[1-]$x_0 \oplus (x_1 \oplus x_2 ) = (x_0 \oplus x_1 ) \oplus x_2 = x_0 \oplus 
x_1 \oplus x_2 $.$(\mbox{Associative)}$

\item[2-]$x_0 (x_1 \oplus x_2 ) = x_0 x_1 \oplus x_0 x_2 $. (Distributive)

\item[3-]$x_0 \oplus x_1 = x_1 \oplus x_0 $. (Commutative)
\end{itemize}

\subsection{Representation of Reed-Muller Expansions}

Any Boolean function $f$ with $n$ variables $f:\left\{ {0,1} \right\}^n \to 
\left\{ {0,1} \right\}$ can be represented as a sum of products \cite{bookRM}:

\begin{equation}
f(x_0 ,...,x_{n - 1} ) = +\sum\limits_{i = 0}^{2^n - 1} {a_i m_i },
\label{eqn1}
\end{equation}

\noindent
where $m_{i}$ are the minterms and $a_{i }$= 0 or 1 indicates the presence or 
absence of minterms respectively and the plus in front of the sigma means that the arguments 
are subject to Boolean operation inclusive-OR. This expansion can also be expressed in (RM) as 
follows \cite{boolth},

\begin{equation}
f({\mathop x\limits^ \bullet}  _0 ,...,{\mathop x\limits^ \bullet}  _{n - 1} ) =  \oplus \sum\limits_{i = 0}^{2^n  - 1} {b_i \varphi _i } 
\label{eqn2}
\end{equation}

\noindent
where

\begin{equation}
\label{eqn3}
\varphi _i = \prod\limits_{k = 0}^{n - 1} {\mathop x\limits^ \bullet}  _{k}^{i_k} 
\end{equation}

\noindent
where ${\mathop x\limits^\bullet} _k = x_k $ or ${\overline x} _k $ and $x_k ,b_i 
\in \left\{ {0,1} \right\}$ and $i_k$ represent the binary digits of $k$.

$\varphi _i$ are known as product terms and $b_i $ determine whether a product 
term is presented or not. $\oplus$ indicates the $XOR$ operation and 
multiplication is assumed to be the $AND$ operation.

A RM function $f({\mathop x\limits^ \bullet}  _0 ,...,{\mathop x\limits^ \bullet}  _{n - 1} )$ 
is said to have {\it fixed polarity} if 
throughout the expansion each variable ${\mathop x\limits^\bullet} _k $ is 
either $x_k$ or ${\overline x}_k $ exclusively. If for some variables $x_k $ 
and ${\overline x} _k $ both occur when the function is said to have 
{\it mixed polarity}.

There is a relation between $a_{i}$ and $b_{i}$ coefficients shown in Eqn.(\ref{eqn1}) 
and Eqn.(\ref{eqn2}), which can be found in detail in \cite{bookRM}.

\subsection {{\bf$\pi$} Notations}

Consider the fixed polarity RM functions with ${\mathop x\limits^\bullet} _k$ 
in its $x_k$ form (Positive Polarity RM). The RM expansion can be expressed 
as a ring sum of products. For $n$ variables expansion, there are $2^{n}$ 
possible combinations of variables known as the $\pi$ terms. 1 and 0 will 
be used to indicate the presence or absence of a variable in the product 
term respectively. For example, a four variable term $x_3 x_2 x_1 x_0$ 
contains the four variables and is represented by 1111 = 15, $x_3 x_2 x_1 
x_0 =\pi _{15} $ and $x_3 x_1 x_0$ ($x_{2}$ is missing) = $\pi _{11}$.

Using this notation \cite{bookRM}, the positive polarity RM expansion shown in Eqn.(\ref{eqn2}) can be 
written as:

\begin{equation}
\label{eqn4}
f(x_0 ,...,x_{n - 1} ) = \oplus \sum\limits_{i = 0}^{2^n - 1} {b_i \pi _i }.
\end{equation}

The conversion between $\varphi _i$ and $\pi _i $ used in Eqn.(\ref{eqn2}) and 
Eqn.(\ref{eqn4}) can be done in both directions. For example, consider the three 
variables $x_{0}$, $x_{1}$ and $x_{2}$:

$
\begin{array}{l}
\varphi _7 = x_0 x_1 x_2 = \pi _7 \\ 
\varphi _6 = x_0 x_1 \overline {x_2 } = x_0 x_1 (x_2 \oplus 1) \\ 
\,\,\,\,\,\,\,\, = x_0 x_1 x_2 \oplus x_0 x_1 \\ 
\,\,\,\,\,\,\,\, = \pi _7 \oplus \pi _6 \\ 
\varphi _5 = x_0 \overline {x_1 } x_2 = x_0 (x_1 \oplus 1)x_2 \\ 
\,\,\,\,\,\,\,\, = x_0 x_1 x_2 \oplus x_0 x_2 \\ 
\,\,\,\,\,\,\,\, = \pi _7 \oplus \pi _5 \\ 
\end{array}
$

Similarly we can construct the rest of conversion as follows:

$
\begin{array}{l}
\varphi _4 = \pi _7 \oplus \pi _6 \oplus \pi _5 \oplus \pi _4 \\
\varphi _3 = \pi _7 \oplus \pi _3 \\ 
\varphi _2 = \pi _7 \oplus \pi _6 \oplus \pi _3 \oplus \pi _2 \\ 
\varphi _1 = \pi _7 \oplus \pi _5 \oplus \pi _3 \oplus \pi _1 \\
\varphi _0 = \pi _7 \oplus \pi _6 \oplus \pi _5 \oplus \pi _4 \oplus \pi _3 
\oplus \pi _2 \oplus \pi _1 \oplus \pi _0 \\
\end{array}
$

\noindent
For the above conversion, the inverse is also true \cite{bookRM},

$
\begin{array}{l}
 \pi _7 = \varphi _7 \\ 
 \pi _6 = \varphi _7 \oplus \varphi _6 \\ 
 \pi _5 = \varphi _7 \oplus \varphi _5 \\ 
 \end{array}
$

\noindent
and so on.

\section{Quantum Boolean Controlled Operations}

\subsection{$CNOT$ Gates}

In our construction for building quantum circuits for Boolean functions, we 
will use one auxiliary qubit, which we initially set to zero, to hold the 
result of the Boolean function; together with $CNOT$ based transformations (Gates) 
which work as follows \cite{transrules}: $CNOT(C \vert t)$ is a gate where the target qubit $t$ is 
controlled by a set of qubits $C$ such that $t \notin C$, the state of the 
qubit $t$ will be flipped from $\left| 0 \right\rangle $ to $\left| 1 
\right\rangle $ or from $\left| 1 \right\rangle $ to $\left| 0 \right\rangle 
$ if and only if all the qubits in $C$ is set to true (state $\left| 1 
\right\rangle )$; i.e. the new state of the target qubit $t$ will be the result 
of $XOR$-ing the old state of $t$ with the $AND$-ing of the states of the control 
qubits. For example, consider the $CNOT$ gate shown in Fig.\ref{fig1}, it can be represented 
as $CNOT\left( {\left\{ {x_0 ,x_2 } 
\right\}\vert x_3 } \right)$, where $\bullet $ on a qubit means that the 
condition on that qubit will evaluate to true if and only if the state of 
that qubit is $\left| 1 \right\rangle $, while $ \oplus $ denotes the target 
qubit which will be flipped if and only if all the control qubits are set 
to true, which means that the state of the qubit $x_{3}$ will be flipped if 
and only if $x_{0}=x_{2}=\left| 1 \right\rangle $ with whatever value 
in $x_{1}$; i.e. $x_{3}$ will be changed according to the operation $x_3 \to 
x_3 \oplus x_0 x_2 $. 

Some special cases of the general $CNOT$ gate have their own names, a $CNOT$ gate with 
one control qubit is called {\it Controlled-Not} gate (Fig.\ref{fig2}-a), $CNOT$ gate with two control qubits 
is called {\it Toffoli} gate (Fig.\ref{fig2}-b) and $CNOT$ gate with no control qubits is called 
{\it NOT} gate (Fig.\ref{fig2}-c) where $C$ will be an empty set (C = $\phi )$, we will refer to this 
case as {\it CNOT}($x_{0})$ where $x_{0}$  is the qubit which will be unconditionally 
flipped.

\begin{center}
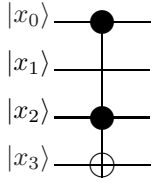
\begin{figure} 
\begin{center}
\setlength{\unitlength}{3947sp}%
\begingroup\makeatletter\ifx\SetFigFont\undefined%
\gdef\SetFigFont#1#2#3#4#5{%
  \reset@font\fontsize{#1}{#2pt}%
  \fontfamily{#3}\fontseries{#4}\fontshape{#5}%
  \selectfont}%
\fi\endgroup%
\begin{picture}(624,1070)(2389,-2548)
{\color[rgb]{0,0,0}\thinlines
\put(2701,-1561){\circle*{150}}
}%
{\color[rgb]{0,0,0}\put(2701,-2161){\circle*{150}}
}%
{\color[rgb]{0,0,0}\put(2701,-2461){\circle{150}}
}%
{\color[rgb]{0,0,0}\put(2401,-1561){\line( 1, 0){600}}
}%
\put(2100,-1561){$\left| {x_0 } \right\rangle$}

{\color[rgb]{0,0,0}\put(2401,-1861){\line( 1, 0){ 75}}
\put(2100,-1861){$\left| {x_1 } \right\rangle$}
\put(2476,-1861){\line( 1, 0){ 75}}
\put(2551,-1861){\line( 1, 0){ 75}}
\put(2626,-1861){\line( 1, 0){ 75}}
\put(2701,-1861){\line( 1, 0){ 75}}
\put(2776,-1861){\line( 1, 0){ 75}}
\put(2851,-1861){\line( 1, 0){ 75}}
\put(2926,-1861){\line( 1, 0){ 75}}
}%
{\color[rgb]{0,0,0}\put(2401,-2161){\line( 1, 0){ 75}}
\put(2100,-2161){$\left| {x_2 } \right\rangle$}

\put(2476,-2161){\line( 1, 0){ 75}}
\put(2551,-2161){\line( 1, 0){ 75}}
\put(2626,-2161){\line( 1, 0){ 75}}
\put(2701,-2161){\line( 1, 0){ 75}}
\put(2776,-2161){\line( 1, 0){ 75}}
\put(2851,-2161){\line( 1, 0){ 75}}
\put(2926,-2161){\line( 1, 0){ 75}}
}%
{\color[rgb]{0,0,0}\put(2401,-2461){\line( 1, 0){600}}
}%
\put(2100,-2461){$\left| {x_3 } \right\rangle$}

{\color[rgb]{0,0,0}\put(2701,-1561){\line( 0,-1){975}}
}%
\end{picture}
\end{center}
\caption{\label{fig1}$CNOT\left( {\left\{ {x_0 ,x_2 } \right\}\vert x_3 } \right)$ gate.}
\end{figure}
\end{center}

\begin{center}
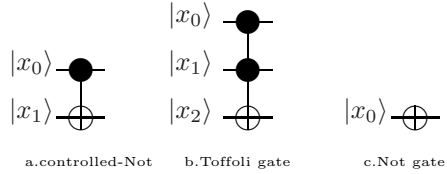
\begin{figure}
\begin{center}
\setlength{\unitlength}{3947sp}%
\begingroup\makeatletter\ifx\SetFigFont\undefined%
\gdef\SetFigFont#1#2#3#4#5{%
  \reset@font\fontsize{#1}{#2pt}%
  \fontfamily{#3}\fontseries{#4}\fontshape{#5}%
  \selectfont}%
\fi\endgroup%
\begin{picture}(2562,1036)(1651,-2492)
{\color[rgb]{0,0,0}\thinlines
\put(4051,-2161){\circle{150}}
}%
{\color[rgb]{0,0,0}\put(1951,-2161){\circle{150}}
}%
{\color[rgb]{0,0,0}\put(1951,-1861){\circle*{150}}
}%
{\color[rgb]{0,0,0}\put(3001,-1561){\circle*{150}}
}%
{\color[rgb]{0,0,0}\put(3001,-1861){\circle*{150}}
}%
{\color[rgb]{0,0,0}\put(3001,-2161){\circle{150}}
}%
{\color[rgb]{0,0,0}\put(3901,-2161){\line( 1, 0){300}}
}%
{\color[rgb]{0,0,0}\put(4051,-2086){\line( 0,-1){150}}
}%
{\color[rgb]{0,0,0}\put(1801,-1861){\line( 1, 0){300}}
}%
{\color[rgb]{0,0,0}\put(1801,-2161){\line( 1, 0){300}}
}%
{\color[rgb]{0,0,0}\put(1951,-1861){\line( 0,-1){375}}
}%
{\color[rgb]{0,0,0}\put(2851,-2161){\line( 1, 0){300}}
}%
{\color[rgb]{0,0,0}\put(2851,-1861){\line( 1, 0){300}}
}%
{\color[rgb]{0,0,0}\put(2851,-1561){\line( 1, 0){300}}
}%
{\color[rgb]{0,0,0}\put(3001,-2236){\line( 0, 1){675}}
}%
\put(3600,-2161){$\left| {x_0 } \right\rangle$}%

\put(1500,-1861){$\left| {x_0 } \right\rangle$}%

\put(1500,-2161){$\left| {x_1 } \right\rangle$}%

\put(2500,-2161){$\left| {x_2 } \right\rangle$}%

\put(2500,-1861){$\left| {x_1 } \right\rangle$}%

\put(2500,-1561){$\left| {x_0 } \right\rangle$}%

\put(1600,-2461){\makebox(0,0)[lb]{\smash{\SetFigFont{5}{6.0}{\rmdefault}{\mddefault}{\updefault}{\color[rgb]{0,0,0}a.controlled-Not}%
}}}
\put(2601,-2461){\makebox(0,0)[lb]{\smash{\SetFigFont{5}{6.0}{\rmdefault}{\mddefault}{\updefault}{\color[rgb]{0,0,0}b.Toffoli gate}%
}}}
\put(3726,-2461){\makebox(0,0)[lb]{\smash{\SetFigFont{5}{6.0}{\rmdefault}{\mddefault}{\updefault}{\color[rgb]{0,0,0}c.Not gate}%
}}}
\end{picture}
\end{center}
\caption{\label{fig2}Special cases of the general $CNOT$ gate.}
\end{figure}
\end{center}

\subsection{Quantum Boolean Circuits}

A general quantum Boolean circuit $U$ of size $m$ over $n$ qubit quantum system 
with qubits $\left| {x_0 } \right\rangle ,\left| {x_1 } \right\rangle 
,\ldots ,\left| {x_{n - 1} } \right\rangle $ can be represented as a 
sequence of \textit{CNOT} gates\cite{transrules},

\begin{equation}
\label{eqn5}
U = CNOT\left( {C_1 \vert t_1 } \right)\ldots CNOT\left( {C_i \vert t_i } 
\right)\ldots CNOT\left( {C_m \vert t_m } \right)
\end{equation}

\noindent
where $t_i \in \left\{ {x_0 ,\ldots ,x_{n - 1} } \right\};\,\,C_i \subset 
\left\{ {x_0 ,\ldots ,x_{n - 1} } \right\};t_i \notin C_i$. The quantum Boolean circuits we will use in this paper can be represented as follows,

\begin{equation}
\label{eqn6}
{U}' = CNOT(C_1 \vert t)...CNOT(C_2 \vert t)...CNOT(C_m \vert t)
\end{equation}

\noindent
where $t \equiv x_{n - 1} ;\,\,C_i \subseteq \left\{ {x_0 ,\ldots ,x_{n - 2} } 
\right\}.$

\begin{center}
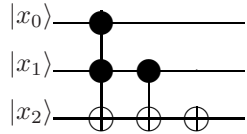
\begin{figure}
\begin{center}
\setlength{\unitlength}{3947sp}%
\begingroup\makeatletter\ifx\SetFigFont\undefined%
\gdef\SetFigFont#1#2#3#4#5{%
  \reset@font\fontsize{#1}{#2pt}%
  \fontfamily{#3}\fontseries{#4}\fontshape{#5}%
  \selectfont}%
\fi\endgroup%
\begin{picture}(1362,792)(1051,-2248)
{\color[rgb]{0,0,0}\thinlines
\put(1501,-1561){\circle*{150}}
}%
{\color[rgb]{0,0,0}\put(1501,-1861){\circle*{150}}
}%
{\color[rgb]{0,0,0}\put(1801,-1861){\circle*{150}}
}%
{\color[rgb]{0,0,0}\put(2101,-2161){\circle{150}}
}%
{\color[rgb]{0,0,0}\put(1801,-2161){\circle{150}}
}%
{\color[rgb]{0,0,0}\put(1501,-2161){\circle{150}}
}%
{\color[rgb]{0,0,0}\put(2101,-1861){\makebox(1.6667,11.6667){\SetFigFont{5}{6}{\rmdefault}{\mddefault}{\updefault}.}}
}%
{\color[rgb]{0,0,0}\put(1201,-1561){\line( 1, 0){1200}}
}%
{\color[rgb]{0,0,0}\put(1201,-1861){\line( 1, 0){1200}}
}%
{\color[rgb]{0,0,0}\put(1201,-2161){\line( 1, 0){1200}}
}%
{\color[rgb]{0,0,0}\put(1501,-2236){\line( 0, 1){750}}
\put(1501,-1486){\line( 0,-1){ 75}}
}%
{\color[rgb]{0,0,0}\put(1801,-2236){\line( 0, 1){375}}
}%
{\color[rgb]{0,0,0}\put(2101,-2236){\line( 0, 1){150}}
}%
\put(920,-1561){$\left| {x_0 } \right\rangle$}%

\put(920,-1861){$\left| {x_1 } \right\rangle$}%

\put(920,-2161){$\left| {x_2 } \right\rangle$}%

\end{picture}
\end{center}
\caption{\label{fig3}Quantum Boolean circuit.}
\end{figure}
\end{center}

For example, consider the quantum circuit shown in Fig.\ref{fig3}, it can be 
represented as follows:

\begin{equation}
\label{eqn7}
U = CNOT(\{x_0 ,x_1 \}\vert x_2 ).CNOT(\{x_1 \}\vert x_2 ).CNOT(x_2 )
\end{equation}

Now, to trace the operations been applied on the target qubit $x_{2}$, we 
shall trace the operation of each of the $CNOT$ gates been applied:

\begin{itemize}
\item $CNOT(\{x_0 ,x_1 \}\vert x_2 ) \Rightarrow x_2 \to x_2 \oplus x_0 x_1$
\item $CNOT(\{x_1 \}\vert x_2 ) \Rightarrow x_2 \to x_2 \oplus x_1$ 
\item $CNOT(x_2 ) \Rightarrow x_2 \to \overline x _2 = x_2 \oplus 1$
\end{itemize}

Combining the three operations, we see that the complete operation applied on 
$x_{2}$ is represented as follows:

\begin{equation}
\label{eqn8}
x_2 \to x_2 \oplus x_0 x_1 \oplus x_1 \oplus 1
\end{equation}

If $x_{2}$ is initialized to $\left| 0 \right\rangle $, applying the 
circuit will make $x_{2}$ carry the result of the operation ($x_0 x_1 \oplus 
x_1 \oplus 1)$, which is equivalent to the operation ($x_0 + \overline x _1 )$.

\section{Representation of Quantum Boolean circuits as RM} 

\subsection{Quantum Boolean Circuits for Positive Polarity RM}

From the above two sections, we may notice that there is a close connection 
between RM and quantum circuits representing arbitrary Boolean function. In 
this section we will show the steps, which we shall follow to implement any 
arbitrary Boolean function $f$ using positive polarity RM expansions as quantum 
circuits.

Example:

Consider the function $f\left( {x_0 ,x_1 ,x_2 } \right) = \overline {x_0 } + 
x_1 x_2 $, to find the quantum circuit implementation for this function; we 
shall follow the following procedure: 

\begin{itemize}
\item[(1)] The above function can be represented as a sum of products as follows:

\begin{equation}
\label{eqn9}
f(x_0 ,x_1 ,x_2 ) = \overline x _0 \overline x _1 \overline x _2 + \overline 
x _0 \overline x _1 x_2 + \overline x _0 x_1 \overline x _2 + \overline x _0 
x_1 x_2 + x_0 x_1 x_2 
\end{equation}

\item[(2)] Converting to $\varphi _i$ notation according to Eqn.(\ref{eqn2}) :

\begin{equation}
\label{eqn11}
f(x_0 ,x_1 ,x_2 ) = \varphi _0 \oplus \varphi _1 \oplus \varphi _2 \oplus 
\varphi _3 \oplus \varphi _7 
\end{equation}

\item[(3)] Substituting $\pi$ product terms shown in section 2.3, we get:

\begin{equation}
\label{eqn12}
\begin{array}{l}
 f = \pi _7 \oplus \pi _6 \oplus \pi _5 \oplus \pi _4 \oplus \pi _3 \oplus 
\pi _2 \oplus \pi _1 \oplus \pi _0 \oplus \pi _7 \oplus \pi _5 \oplus \\ 
 \,\,\,\,\,\,\,\,\,\pi _3 \oplus \pi _1 \oplus \pi _7 \oplus \pi _6 \oplus 
\pi _3 \oplus \pi _2 \oplus \pi _7 \oplus \pi _3 \oplus \pi _7 \\ 
 \end{array}
\end{equation}

\item[(4)] Using modulo-2 operations to simplify this expansion we get,

\begin{equation}
\label{eqn13}
f = \pi _7 \oplus \pi _4 \oplus \pi _0 = x_0 x_1 x_2 \oplus x_0 \oplus 1
\end{equation}

\end{itemize}

Using the last expansion in Eqn.(\ref{eqn13}), we can create the quantum circuit, 
which implements this function as follows:

\medskip

\begin{itemize}

\item[1-]Initialize the target qubit to the state $\left| 0 \right\rangle $, which 
will hold the result of the Boolean function.

\item[2-]Add $CNOT$ gate for each product term in this expansion taking the Boolean 
variables in this product term as control qubits and the result qubit as the 
target qubit $t$.

\item[3-]For the product term, which contains 1, we will add $CNOT(t)$, so the final 
circuit will be as shown in Fig.\ref{fig4}.
\end{itemize}

\begin{center}
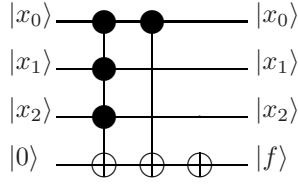
\begin{figure}
\begin{center}
\setlength{\unitlength}{3947sp}%
\begingroup\makeatletter\ifx\SetFigFont\undefined%
\gdef\SetFigFont#1#2#3#4#5{%
  \reset@font\fontsize{#1}{#2pt}%
  \fontfamily{#3}\fontseries{#4}\fontshape{#5}%
  \selectfont}%
\fi\endgroup%
\begin{picture}(1425,1092)(1051,-2248)
{\color[rgb]{0,0,0}\thinlines
\put(1501,-1561){\circle*{150}}
}%
{\color[rgb]{0,0,0}\put(1501,-1861){\circle*{150}}
}%
{\color[rgb]{0,0,0}\put(2101,-2161){\circle{150}}
}%
{\color[rgb]{0,0,0}\put(1801,-2161){\circle{150}}
}%
{\color[rgb]{0,0,0}\put(1501,-2161){\circle{150}}
}%
{\color[rgb]{0,0,0}\put(1501,-1261){\circle*{150}}
}%
{\color[rgb]{0,0,0}\put(1801,-1261){\circle*{150}}
}%
{\color[rgb]{0,0,0}\put(2101,-1861){\makebox(1.6667,11.6667){\SetFigFont{5}{6}{\rmdefault}{\mddefault}{\updefault}.}}
}%
{\color[rgb]{0,0,0}\put(1201,-1561){\line( 1, 0){1200}}
}%
{\color[rgb]{0,0,0}\put(1201,-1861){\line( 1, 0){1200}}
}%
{\color[rgb]{0,0,0}\put(1201,-2161){\line( 1, 0){1200}}
}%
{\color[rgb]{0,0,0}\put(2101,-2236){\line( 0, 1){150}}
}%
{\color[rgb]{0,0,0}\put(1201,-1261){\line( 1, 0){1200}}
}%
{\color[rgb]{0,0,0}\put(1801,-2236){\line( 0, 1){975}}
}%
{\color[rgb]{0,0,0}\put(1501,-2236){\line( 0, 1){975}}
}%
\put(900,-1261){$\left| {x_0 } \right\rangle$}%

\put(900,-1561){$\left| {x_1 } \right\rangle$}%

\put(900,-1861){$\left| {x_2 } \right\rangle$}%

\put(2450,-1261){$\left| {x_0 } \right\rangle$}%

\put(2450,-1561){$\left| {x_1 } \right\rangle$}%

\put(2450,-1861){$\left| {x_2 } \right\rangle$}%

\put(900,-2161){$\left| {0 } \right\rangle$}%

\put(2450,-2161){$\left| {f } \right\rangle$}%

\end{picture}

\end{center}
\caption{\label{fig4}Quantum circuit implementation for $f\left({x_0 ,x_1 ,x_2 } \right) = \overline {x_0} + x_1 x_2$.}
\end{figure}
\end{center}

\subsection{Quantum Circuits for Different RM Polarities }

Consider RM expansion shown in Eqn.(\ref{eqn2}) where ${\mathop x\limits^\bullet}_k$ can be 
$x_k$ or $\overline x _k$ exclusively. For $n$ variables expansion where each 
variable may be in its true or complemented form, but not both, then there 
will be $2^{n}$ possible expansions. 
These are known as {\it fixed polarity generalized Reed-Muller} (GRM) {\it expansions}.

We can identify different GRM expansions by a {\it polarity number}, which is a 
number that represents the binary number calculated in the following way: If a variable 
appears in its true form, it will be represented by 1, and 0 for a variable 
appearing in its complemented form. For example, consider the Boolean function 
$f({\mathop x\limits^\bullet} _0 ,{\mathop x\limits^\bullet} _1 ,{\mathop 
x\limits^\bullet} _2 )$: $f(x_0 ,x_1 ,x_2 )$ has polarity 0 (000), $f(x_0 
,\overline {x_1 } ,x_2 )$ has polarity 2 (010), $f(\overline {x_0 } ,x_1 
,\overline {x_2 } )$ has polarity 5 (101) and $f(\overline {x_0 } ,\overline 
{x_1 } ,\overline {x_2 } )$ has polarity 7 (111) and so on.

RM expansion with a certain polarity can be converted to another polarity by 
replacing any variable $x_i $ by $(\overline {x_i } \oplus 1)$ or any 
variable $\overline {x_i } $ by $(x_i \oplus 1)$. For example, consider the 
Boolean function $f\left( {x_0 ,x_1 ,x_2 } \right) = \overline {x_0 } + x_1 
x_2 $, it can be represented with different polarity RM expansions as 
follows:

\begin{equation}
\label{eqn14}
f = x_0 x_1 x_2 \oplus x_0 \oplus 1 : \mbox{0 polarity}.
\end{equation}

\begin{equation}
\label{eqn15}
f = x_0 x_1 \overline {x_2 } \oplus x_0 x_1 \oplus x_0 \oplus 1 : \mbox{1 polarity}.
\end{equation}

\begin{equation}
\label{eqn16}
f = \overline {x_0 } x_1 \overline {x_2 } \oplus x_1 \overline {x_2 } \oplus 
\overline {x_0 } x_1 \oplus x_1 \oplus \overline {x_0 } : \mbox{5 polarity}.
\end{equation}

\begin{equation}
\label{eqn17}
f = \overline {x_0 }\,\overline {x_1 }\,\overline {x_2 } \oplus \overline {x_0 
}\,\overline {x_2 } \oplus \overline {x_1 }\,\overline {x_2 } \oplus \overline 
{x_0 }\,\overline {x_1 } \oplus \overline {x_1 } \oplus \overline {x_2 } 
\oplus 1 : \mbox{7 polarity}.
\end{equation}

Different polarity RM expansions will give different quantum 
circuits for the same Boolean function. For example, consider different 
polarity representations for the function $f\left( {x_0 ,x_1 ,x_2 } \right) 
= \overline {x_0 } + x_1 x_2 $ shown above. Each representation has 
different quantum circuit (as shown in Fig.\ref{fig5}) using the following procedure:

\medskip

\begin{itemize}
\item[1-]Initialize the target qubit to the state $\left| 0 \right\rangle$, which 
will hold the result of the Boolean function.

\item[2-]Add $CNOT$ gate for each product term in the RM expansion taking the Boolean 
variables in this term as control qubits and the result qubit as the target 
qubit $t$.

\item[3-]For the product term, which contains 1, we will add $CNOT(t)$.

\item[4-]For control qubit $x_{i}$, which appears in complemented form, we will add 
$CNOT(x_{i})$ at the beginning of the circuit to negate it's value during the 
run of the circuit and add another $CNOT(x_{i})$ at the end of the circuit to 
restore it's original value.
\end{itemize}

It is clear from Fig.\ref{fig5} that changing polarity will change the number of $CNOT$ 
gates in the circuits; i.e. its efficiency. This means that there is a need to develop search 
algorithms for optimizing canonical Reed- Muller expansions for quantum 
Boolean functions similar to those found for classical digital circuit 
design \cite{miller1,miller2}, taking into account that efficient expansions for classical computers 
may be not so efficient for quantum computers. For example, consider $f(x_0 ,x_1 ,x_2 )$ defined as 
follows: 

\begin{equation}
\label{eqn18}
f = \overline x _0 \overline x _1 \overline x _2 + \overline x _0 x_1 
\overline x _2 + x_0 \overline x _1 x_2 + x_0 x_1 x_2, 
\end{equation}

its 0 polarity expansion is given by $(x_0 \oplus x_2 \oplus 1)$ and its 3 
polarity expansion is given by $(\overline x _0 \oplus x_2 )$. From 
a classical point of view, 3 polarity expansion is better than 0 polarity 
expansion because it contains two product terms rather that three product 
terms in 0 polarity expansion. On contrary, on implementing both expansions 
as quantum Boolean circuits we can see that 0 polarity expansion is better 
than 3 polarity expansion because of the number of $CNOT$ gates used as shown 
in Fig.\ref{fig6}.

\begin{center}
\begin{figure}
\begin{center}
\setlength{\unitlength}{3947sp}%
\begingroup\makeatletter\ifx\SetFigFont\undefined%
\gdef\SetFigFont#1#2#3#4#5{%
  \reset@font\fontsize{#1}{#2pt}%
  \fontfamily{#3}\fontseries{#4}\fontshape{#5}%
  \selectfont}%
\fi\endgroup%
\begin{picture}(4725,2836)(751,-3992)
{\color[rgb]{0,0,0}\thinlines
\put(1501,-1561){\circle*{150}}
}%
{\color[rgb]{0,0,0}\put(1501,-1861){\circle*{150}}
}%
{\color[rgb]{0,0,0}\put(2101,-2161){\circle{150}}
}%
{\color[rgb]{0,0,0}\put(1801,-2161){\circle{150}}
}%
{\color[rgb]{0,0,0}\put(1501,-2161){\circle{150}}
}%
{\color[rgb]{0,0,0}\put(1501,-1261){\circle*{150}}
}%
{\color[rgb]{0,0,0}\put(1801,-1261){\circle*{150}}
}%
{\color[rgb]{0,0,0}\put(3451,-1861){\circle{150}}
}%
{\color[rgb]{0,0,0}\put(3676,-1261){\circle*{150}}
}%
{\color[rgb]{0,0,0}\put(3676,-1561){\circle*{150}}
}%
{\color[rgb]{0,0,0}\put(3676,-1861){\circle*{150}}
}%
{\color[rgb]{0,0,0}\put(3676,-2161){\circle{150}}
}%
{\color[rgb]{0,0,0}\put(3976,-1261){\circle*{150}}
}%
{\color[rgb]{0,0,0}\put(3976,-2161){\circle{150}}
}%
{\color[rgb]{0,0,0}\put(4276,-1261){\circle*{150}}
}%
{\color[rgb]{0,0,0}\put(4276,-1561){\circle*{150}}
}%
{\color[rgb]{0,0,0}\put(4276,-2161){\circle{150}}
}%
{\color[rgb]{0,0,0}\put(4576,-1861){\circle{150}}
}%
{\color[rgb]{0,0,0}\put(4576,-2161){\circle{150}}
}%
{\color[rgb]{0,0,0}\put(1051,-2761){\circle{150}}
}%
{\color[rgb]{0,0,0}\put(1051,-3361){\circle{150}}
}%
{\color[rgb]{0,0,0}\put(1276,-2761){\circle*{150}}
}%
{\color[rgb]{0,0,0}\put(1276,-3061){\circle*{150}}
}%
{\color[rgb]{0,0,0}\put(1276,-3361){\circle*{150}}
}%
{\color[rgb]{0,0,0}\put(1501,-3061){\circle*{150}}
}%
{\color[rgb]{0,0,0}\put(1501,-3361){\circle*{150}}
}%
{\color[rgb]{0,0,0}\put(1726,-2761){\circle*{150}}
}%
{\color[rgb]{0,0,0}\put(1726,-3061){\circle*{150}}
}%
{\color[rgb]{0,0,0}\put(1951,-3061){\circle*{150}}
}%
{\color[rgb]{0,0,0}\put(2476,-2761){\circle{150}}
}%
{\color[rgb]{0,0,0}\put(2476,-3361){\circle{150}}
}%
{\color[rgb]{0,0,0}\put(2251,-2761){\circle*{150}}
}%
{\color[rgb]{0,0,0}\put(1276,-3661){\circle{150}}
}%
{\color[rgb]{0,0,0}\put(1501,-3661){\circle{150}}
}%
{\color[rgb]{0,0,0}\put(1726,-3661){\circle{150}}
}%
{\color[rgb]{0,0,0}\put(1951,-3661){\circle{150}}
}%
{\color[rgb]{0,0,0}\put(2251,-3661){\circle{150}}
}%
{\color[rgb]{0,0,0}\put(3451,-3061){\circle{150}}
}%
{\color[rgb]{0,0,0}\put(3451,-3361){\circle{150}}
}%
{\color[rgb]{0,0,0}\put(3676,-3061){\circle*{150}}
}%
{\color[rgb]{0,0,0}\put(3676,-3361){\circle*{150}}
}%
{\color[rgb]{0,0,0}\put(3451,-2761){\circle{150}}
}%
{\color[rgb]{0,0,0}\put(3676,-2761){\circle*{150}}
}%
{\color[rgb]{0,0,0}\put(3901,-2761){\circle*{150}}
}%
{\color[rgb]{0,0,0}\put(3901,-3361){\circle*{150}}
}%
{\color[rgb]{0,0,0}\put(4126,-3061){\circle*{150}}
}%
{\color[rgb]{0,0,0}\put(4126,-3361){\circle*{150}}
}%
{\color[rgb]{0,0,0}\put(4351,-2761){\circle*{150}}
}%
{\color[rgb]{0,0,0}\put(4351,-3061){\circle*{150}}
}%
{\color[rgb]{0,0,0}\put(4576,-3061){\circle*{150}}
}%
{\color[rgb]{0,0,0}\put(4801,-3361){\circle*{150}}
}%
{\color[rgb]{0,0,0}\put(5101,-2761){\circle{150}}
}%
{\color[rgb]{0,0,0}\put(5101,-3061){\circle{150}}
}%
{\color[rgb]{0,0,0}\put(5101,-3361){\circle{150}}
}%
{\color[rgb]{0,0,0}\put(5101,-3661){\circle{150}}
}%
{\color[rgb]{0,0,0}\put(4801,-3661){\circle{150}}
}%
{\color[rgb]{0,0,0}\put(4576,-3661){\circle{150}}
}%
{\color[rgb]{0,0,0}\put(4351,-3661){\circle{150}}
}%
{\color[rgb]{0,0,0}\put(4126,-3661){\circle{150}}
}%
{\color[rgb]{0,0,0}\put(3901,-3661){\circle{150}}
}%
{\color[rgb]{0,0,0}\put(3676,-3661){\circle{150}}
}%
{\color[rgb]{0,0,0}\put(1201,-1561){\line( 1, 0){1200}}
}%
{\color[rgb]{0,0,0}\put(1201,-1861){\line( 1, 0){1200}}
}%
{\color[rgb]{0,0,0}\put(1201,-2161){\line( 1, 0){1200}}
}%
{\color[rgb]{0,0,0}\put(2101,-2236){\line( 0, 1){150}}
}%
{\color[rgb]{0,0,0}\put(1201,-1261){\line( 1, 0){1200}}
}%
{\color[rgb]{0,0,0}\put(1801,-2236){\line( 0, 1){975}}
}%
{\color[rgb]{0,0,0}\put(1501,-2236){\line( 0, 1){975}}
}%
{\color[rgb]{0,0,0}\put(3301,-2161){\line( 1, 0){1575}}
}%
{\color[rgb]{0,0,0}\put(3301,-1861){\line( 1, 0){1575}}
}%
{\color[rgb]{0,0,0}\put(3301,-1561){\line( 1, 0){1575}}
}%
{\color[rgb]{0,0,0}\put(3301,-1261){\line( 1, 0){1575}}
}%
{\color[rgb]{0,0,0}\put(3451,-1786){\line( 0,-1){150}}
}%
{\color[rgb]{0,0,0}\put(3676,-1261){\line( 0,-1){975}}
}%
{\color[rgb]{0,0,0}\put(3976,-1261){\line( 0,-1){975}}
}%
{\color[rgb]{0,0,0}\put(4276,-1261){\line( 0,-1){975}}
}%
{\color[rgb]{0,0,0}\put(4576,-1786){\line( 0,-1){150}}
}%
{\color[rgb]{0,0,0}\put(4576,-2086){\line( 0,-1){150}}
}%
{\color[rgb]{0,0,0}\put(901,-2761){\line( 1, 0){1800}}
}%
{\color[rgb]{0,0,0}\put(901,-3061){\line( 1, 0){1800}}
}%
{\color[rgb]{0,0,0}\put(901,-3361){\line( 1, 0){1800}}
}%
{\color[rgb]{0,0,0}\put(901,-3661){\line( 1, 0){1800}}
}%
{\color[rgb]{0,0,0}\put(1276,-2761){\line( 0,-1){975}}
}%
{\color[rgb]{0,0,0}\put(1501,-3061){\line( 0,-1){675}}
}%
{\color[rgb]{0,0,0}\put(1726,-2761){\line( 0,-1){975}}
}%
{\color[rgb]{0,0,0}\put(1951,-3061){\line( 0,-1){675}}
}%
{\color[rgb]{0,0,0}\put(2251,-2761){\line( 0,-1){975}}
}%
{\color[rgb]{0,0,0}\put(2476,-2686){\line( 0,-1){150}}
}%
{\color[rgb]{0,0,0}\put(2476,-3286){\line( 0,-1){150}}
}%
{\color[rgb]{0,0,0}\put(1051,-2686){\line( 0,-1){150}}
}%
{\color[rgb]{0,0,0}\put(1051,-3286){\line( 0,-1){150}}
}%
{\color[rgb]{0,0,0}\put(3301,-3061){\line( 1, 0){2100}}
}%
{\color[rgb]{0,0,0}\put(3301,-3361){\line( 1, 0){2100}}
}%
{\color[rgb]{0,0,0}\put(3301,-3661){\line( 1, 0){2100}}
}%
{\color[rgb]{0,0,0}\put(3301,-2761){\line( 1, 0){2100}}
}%
{\color[rgb]{0,0,0}\put(3451,-2686){\line( 0,-1){150}}
}%
{\color[rgb]{0,0,0}\put(3451,-2986){\line( 0,-1){150}}
}%
{\color[rgb]{0,0,0}\put(3451,-3286){\line( 0,-1){150}}
}%
{\color[rgb]{0,0,0}\put(3676,-2761){\line( 0,-1){975}}
}%
{\color[rgb]{0,0,0}\put(3901,-2761){\line( 0,-1){975}}
}%
{\color[rgb]{0,0,0}\put(4126,-3061){\line( 0,-1){675}}
}%
{\color[rgb]{0,0,0}\put(4351,-2761){\line( 0,-1){975}}
}%
{\color[rgb]{0,0,0}\put(4576,-3061){\line( 0,-1){675}}
}%
{\color[rgb]{0,0,0}\put(4801,-3361){\line( 0,-1){375}}
}%
{\color[rgb]{0,0,0}\put(5101,-2686){\line( 0,-1){150}}
}%
{\color[rgb]{0,0,0}\put(5101,-2986){\line( 0,-1){150}}
}%
{\color[rgb]{0,0,0}\put(5101,-3286){\line( 0,-1){150}}
}%
{\color[rgb]{0,0,0}\put(5101,-3586){\line( 0,-1){150}}
}%
\put(920,-1261){$\left| {x_0 } \right\rangle$}%

\put(920,-1561){$\left| {x_1 } \right\rangle$}%

\put(920,-1861){$\left| {x_2 } \right\rangle$}%

\put(2476,-1261){$\left| {x_0} \right\rangle$}%

\put(2476,-1561){$\left| {x_1 } \right\rangle$}%

\put(2476,-1861){$\left| {x_2 } \right\rangle$}%

\put(920,-2161){$\left| {0 } \right\rangle$}%

\put(2476,-2161){$\left| {f} \right\rangle$}%

\put(2920,-1261){$\left| {x_0} \right\rangle$}%

\put(2920,-1561){$\left| {x_1 } \right\rangle$}%

\put(2920,-1861){$\left| {x_2 } \right\rangle$}%

\put(2920,-2161){$\left| {0 } \right\rangle$}%

\put(4951,-1261){$\left| {x_0 } \right\rangle$}%

\put(4951,-1561){$\left| {x_1 } \right\rangle$}%

\put(4951,-1861){$\left| {x_2 } \right\rangle$}%

\put(4951,-2161){$\left| {f } \right\rangle$}%

\put(1576,-2461){\makebox(0,0)[lb]{\smash{\SetFigFont{10}{6.0}{\rmdefault}{\mddefault}{\updefault}{\color[rgb]{0,0,0}0 Polarity}%
}}}
\put(3976,-2461){\makebox(0,0)[lb]{\smash{\SetFigFont{10}{6.0}{\rmdefault}{\mddefault}{\updefault}{\color[rgb]{0,0,0}1 Polarity}%
}}}
\put(651,-2761){$\left| {x_0 } \right\rangle$}%

\put(2700,-2761){$\left| {x_0 } \right\rangle$}%

\put(651,-3061){$\left| {x_1 } \right\rangle$}%

\put(2700,-3061){$\left| {x_1 } \right\rangle$}%

\put(651,-3361){$\left| {x_2 } \right\rangle$}%

\put(2700,-3361){$\left| {x_2 } \right\rangle$}%

\put(651,-3661){$\left| {0 } \right\rangle$}%

\put(2700,-3661){$\left| {f } \right\rangle$}%

\put(3051,-2761){$\left| {x_0 } \right\rangle$}%

\put(3051,-3061){$\left| {x_1 } \right\rangle$}%

\put(3051,-3361){$\left| {x_2 } \right\rangle$}%

\put(5400,-2761){$\left| {x_0 } \right\rangle$}%

\put(5400,-3061){$\left| {x_1 } \right\rangle$}%

\put(5400,-3361){$\left| {x_2 } \right\rangle$}%

\put(3051,-3661){$\left| {0 } \right\rangle$}%

\put(5400,-3661){$\left| {f } \right\rangle$}%

\put(1576,-3961){\makebox(0,0)[lb]{\smash{\SetFigFont{10}{6.0}{\rmdefault}{\mddefault}{\updefault}{\color[rgb]{0,0,0}5 Polarity}%
}}}
\put(3976,-3961){\makebox(0,0)[lb]{\smash{\SetFigFont{10}{6.0}{\rmdefault}{\mddefault}{\updefault}{\color[rgb]{0,0,0}7 Polarity}%
}}}
\end{picture}

\end{center}
\caption{\label{fig5}Quantum circuits for the Boolean function $f\left( {x_0 ,x_1 ,x_2 
} \right) = \overline {x_0 } + x_1 x_2 $ with different Polarities.}
\end{figure}
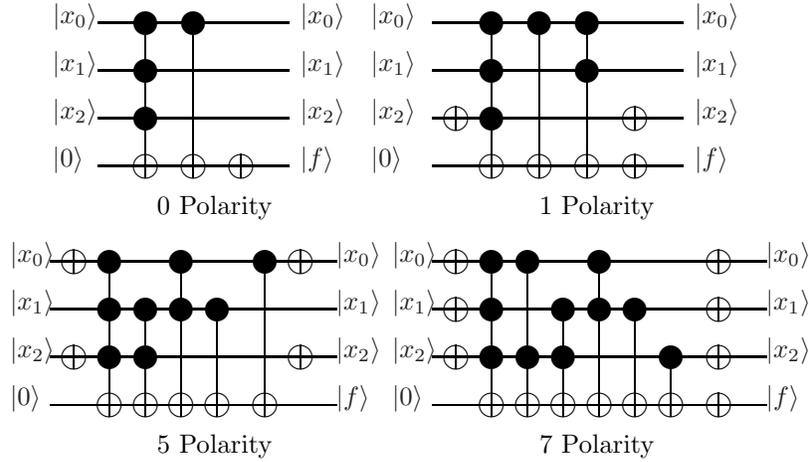
\end{center}

\begin{center}
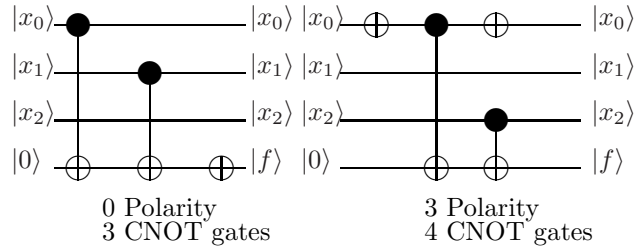
\begin{figure}
\begin{center}

\setlength{\unitlength}{3947sp}%
\begingroup\makeatletter\ifx\SetFigFont\undefined%
\gdef\SetFigFont#1#2#3#4#5{%
  \reset@font\fontsize{#1}{#2pt}%
  \fontfamily{#3}\fontseries{#4}\fontshape{#5}%
  \selectfont}%
\fi\endgroup%
\begin{picture}(3525,1495)(1051,-2651)
{\color[rgb]{0,0,0}\thinlines
\put(1351,-1261){\circle*{150}}
}%
{\color[rgb]{0,0,0}\put(1351,-2161){\circle{150}}
}%
{\color[rgb]{0,0,0}\put(1801,-1561){\circle*{150}}
}%
{\color[rgb]{0,0,0}\put(1801,-2161){\circle{150}}
}%
{\color[rgb]{0,0,0}\put(2251,-2161){\circle{150}}
}%
{\color[rgb]{0,0,0}\put(3226,-1261){\circle{150}}
}%
{\color[rgb]{0,0,0}\put(3601,-1261){\circle*{150}}
}%
{\color[rgb]{0,0,0}\put(3601,-2161){\circle{150}}
}%
{\color[rgb]{0,0,0}\put(3976,-1261){\circle{150}}
}%
{\color[rgb]{0,0,0}\put(3976,-1861){\circle*{150}}
}%
{\color[rgb]{0,0,0}\put(3976,-2161){\circle{150}}
}%
{\color[rgb]{0,0,0}\put(1201,-1561){\line( 1, 0){1200}}
}%
{\color[rgb]{0,0,0}\put(1201,-1861){\line( 1, 0){1200}}
}%
{\color[rgb]{0,0,0}\put(1201,-2161){\line( 1, 0){1200}}
}%
{\color[rgb]{0,0,0}\put(1201,-1261){\line( 1, 0){1200}}
}%
{\color[rgb]{0,0,0}\put(3001,-1261){\line( 1, 0){1500}}
}%
{\color[rgb]{0,0,0}\put(3001,-1561){\line( 1, 0){1500}}
}%
{\color[rgb]{0,0,0}\put(3001,-1861){\line( 1, 0){1500}}
}%
{\color[rgb]{0,0,0}\put(3001,-2161){\line( 1, 0){1500}}
}%
{\color[rgb]{0,0,0}\put(1351,-1261){\line( 0,-1){975}}
}%
{\color[rgb]{0,0,0}\put(1801,-1561){\line( 0,-1){675}}
}%
{\color[rgb]{0,0,0}\put(2251,-2086){\line( 0,-1){150}}
}%
{\color[rgb]{0,0,0}\put(3226,-1186){\line( 0,-1){150}}
}%
{\color[rgb]{0,0,0}\put(3601,-1261){\line( 0,-1){975}}
}%
{\color[rgb]{0,0,0}\put(3976,-1186){\line( 0,-1){150}}
}%
{\color[rgb]{0,0,0}\put(3976,-1861){\line( 0,-1){375}}
}%
\put(930,-1261){$\left| {x_0 } \right\rangle$}%

\put(930,-1561){$\left| {x_1 } \right\rangle$}%

\put(930,-1861){$\left| {x_2 } \right\rangle$}%

\put(2430,-1261){$\left| {x_0 } \right\rangle$}%

\put(2430,-1561){$\left| {x_1 } \right\rangle$}%

\put(2430,-1861){$\left| {x_2 } \right\rangle$}%

\put(930,-2161){$\left| {0 } \right\rangle$}%

\put(2430,-2161){$\left| {f } \right\rangle$}%

\put(2750,-1261){$\left| {x_0 } \right\rangle$}%

\put(2750,-1561){$\left| {x_1 } \right\rangle$}%

\put(2750,-1861){$\left| {x_2 } \right\rangle$}%

\put(4576,-2161){$\left| {f } \right\rangle$}%

\put(4576,-1861){$\left| {x_2 } \right\rangle$}%

\put(4576,-1561){$\left| {x_1 } \right\rangle$}%

\put(4576,-1261){$\left| {x_0 } \right\rangle$}%

\put(2750,-2161){$\left| {0 } \right\rangle$}%

\put(1501,-2461){\makebox(0,0)[lb]{\smash{\SetFigFont{10}{12.0}{\rmdefault}{\mddefault}{\updefault}{\color[rgb]{0,0,0}0 Polarity}%
}}}
\put(1501,-2611){\makebox(0,0)[lb]{\smash{\SetFigFont{10}{12.0}{\rmdefault}{\mddefault}{\updefault}{\color[rgb]{0,0,0}3 CNOT gates}%
}}}
\put(3526,-2461){\makebox(0,0)[lb]{\smash{\SetFigFont{10}{12.0}{\rmdefault}{\mddefault}{\updefault}{\color[rgb]{0,0,0}3 Polarity}%
}}}
\put(3526,-2611){\makebox(0,0)[lb]{\smash{\SetFigFont{10}{12.0}{\rmdefault}{\mddefault}{\updefault}{\color[rgb]{0,0,0}4 CNOT gates}%
}}}
\end{picture}
\end{center}
\caption{\label{fig6}Changing polarity may affect the number of $CNOT$ gates used.}
\end{figure}
\end{center}

\subsection{Boolean Quantum Circuits for Mixed Polarity RM}

Mixed polarity RM are expansions where it is allowed for some variables 
${\mathop x\limits^\bullet} _k$ to appear in its true form $(x_k )$ and 
complemented form $(\overline x _k)$ both in the same expansion. To 
understand how this kind of expansions can be implemented as a quantum 
circuit, consider the three variable mixed polarity RM,

\begin{equation}
\label{19}
f = \overline {x_0} x_1 x_2 \oplus x_0 \overline {x_1 } \oplus x_0 \oplus \overline {x_2 } \oplus 1,
\end{equation}

\noindent 
using the following procedure, we will get the quantum circuit as 
shown in Fig.\ref{fig7}:

\medskip

\begin{itemize}
\item[1-]Initialize the target qubit to the state $\left| 0 \right\rangle $, which 
will hold the result of the Boolean function.

\item[2-]Add $CNOT$ gate for each product term in the RM expansion taking the Boolean 
variables in this term as control qubits and the result qubit as the target 
qubit $t$.

\item[3-]For the product term, which contains 1, we will add $CNOT(t)$.

\item[4-]For control qubit $x_{i}$, which appears in complemented form, we will add 
$CNOT(x_{i})$ directly before and after (negate/restore) the $CNOT$ gate where this 
variable appears in its complemented form.

\end{itemize}

\begin{center}
\begin{figure}
\begin{center}

\setlength{\unitlength}{3947sp}%
\begingroup\makeatletter\ifx\SetFigFont\undefined%
\gdef\SetFigFont#1#2#3#4#5{%
  \reset@font\fontsize{#1}{#2pt}%
  \fontfamily{#3}\fontseries{#4}\fontshape{#5}%
  \selectfont}%
\fi\endgroup%
\begin{picture}(3000,1092)(1051,-2248)
{\color[rgb]{0,0,0}\thinlines
\put(1426,-1261){\circle{150}}
}%
{\color[rgb]{0,0,0}\put(1726,-1261){\circle*{150}}
}%
{\color[rgb]{0,0,0}\put(1726,-1561){\circle*{150}}
}%
{\color[rgb]{0,0,0}\put(1726,-1861){\circle*{150}}
}%
{\color[rgb]{0,0,0}\put(1726,-2161){\circle{150}}
}%
{\color[rgb]{0,0,0}\put(2101,-1561){\circle{150}}
}%
{\color[rgb]{0,0,0}\put(2401,-1261){\circle*{150}}
}%
{\color[rgb]{0,0,0}\put(2401,-1561){\circle*{150}}
}%
{\color[rgb]{0,0,0}\put(2401,-2161){\circle{150}}
}%
{\color[rgb]{0,0,0}\put(2701,-1561){\circle{150}}
}%
{\color[rgb]{0,0,0}\put(3076,-1261){\circle*{150}}
}%
{\color[rgb]{0,0,0}\put(3076,-2161){\circle{150}}
}%
{\color[rgb]{0,0,0}\put(3301,-1861){\circle{150}}
}%
{\color[rgb]{0,0,0}\put(3526,-1861){\circle*{150}}
}%
{\color[rgb]{0,0,0}\put(3751,-1861){\circle{150}}
}%
{\color[rgb]{0,0,0}\put(3526,-2161){\circle{150}}
}%
{\color[rgb]{0,0,0}\put(3751,-2161){\circle{150}}
}%
{\color[rgb]{0,0,0}\put(2101,-1261){\circle{150}}
}%
{\color[rgb]{0,0,0}\put(1201,-1261){\line( 1, 0){2700}}
}%
{\color[rgb]{0,0,0}\put(1201,-1561){\line( 1, 0){2700}}
}%
{\color[rgb]{0,0,0}\put(1201,-1861){\line( 1, 0){2700}}
}%
{\color[rgb]{0,0,0}\put(1201,-2161){\line( 1, 0){2700}}
}%
{\color[rgb]{0,0,0}\put(1426,-1186){\line( 0,-1){150}}
}%
{\color[rgb]{0,0,0}\put(1726,-1261){\line( 0,-1){975}}
}%
{\color[rgb]{0,0,0}\put(2101,-1486){\line( 0,-1){150}}
}%
{\color[rgb]{0,0,0}\put(2401,-1261){\line( 0,-1){975}}
}%
{\color[rgb]{0,0,0}\put(2701,-1486){\line( 0,-1){150}}
}%
{\color[rgb]{0,0,0}\put(3076,-1261){\line( 0,-1){975}}
}%
{\color[rgb]{0,0,0}\put(3301,-1786){\line( 0,-1){150}}
}%
{\color[rgb]{0,0,0}\put(3526,-1861){\line( 0,-1){375}}
}%
{\color[rgb]{0,0,0}\put(3751,-1786){\line( 0,-1){150}}
}%
{\color[rgb]{0,0,0}\put(3751,-2086){\line( 0,-1){150}}
}%
{\color[rgb]{0,0,0}\put(2101,-1186){\line( 0,-1){150}}
}%

\put(920,-1261){$\left| {x_0 } \right\rangle$}%

\put(920,-1561){$\left| {x_1 } \right\rangle$}%

\put(920,-1861){$\left| {x_2 } \right\rangle$}%

\put(920,-2161){$\left| {0 } \right\rangle$}%

\put(4051,-1261){$\left| {x_0 } \right\rangle$}%

\put(4051,-1561){$\left| {x_1 } \right\rangle$}%

\put(4051,-1861){$\left| {x_2 } \right\rangle$}%

\put(4051,-2161){$\left| {f } \right\rangle$}%

\end{picture}

\end{center}
\caption{\label{fig7}Mixed polarity quantum Boolean circuit for$f = \overline {x_0 } 
x_1 x_2 \oplus x_0 \overline {x_1 } \oplus x_0 \oplus \overline {x_2 } 
\oplus 1$.}
\end{figure}
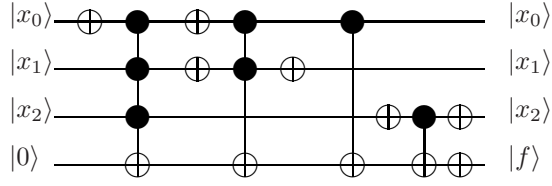
\end{center}

\subsection{Calculating Total Number of $CNOT$ gates}

For {\bf Fixed Polarity} RM expansion, the number of $CNOT$ gates in the final 
quantum circuit can be calculated as follows:

\begin{equation}
\label{eqn20}
S_1 = m + 2K,
\quad
0 \le m \le 2^n;\,0 \le K \le n,
\end{equation}

\noindent
where $S_{1}$ is the total number of $CNOT$ gates, $m$ is the number of product terms 
in the expansion, $K$ is the number of variables in complemented form and $n$ is 
the number of inputs to the Boolean function, the term $2K$ represents the number 
of $CNOT$ gates which will be added at the beginning and the ending of the 
circuit (complemented form) to negate the value of the control qubit during 
the run of the circuit and to restore it's original value respectively.

For {\bf Mixed Polarity} RM expansion, the number of $CNOT$ gates in the final 
quantum circuit can be calculated as follows:

\begin{equation}
\label{eqm21}
S_2 = m + 2L,
\quad
0 \le m \le 2^n;\,1 \le L \le n2^{n - 1}
\end{equation}

\noindent
where $S_{2}$ is the total number of $CNOT$ gates, $m$ is the number of product terms 
in the expansion, $L$ is the total number of occurrences of all variables in 
complemented form and $n$ is the number of inputs to the Boolean function, 
the term $2L$ represents the number of CNOT gates which may be added 
before and after the control qubit which appears in complemented form during 
the run of the circuit to negate/restore it's value respectively.

\section{Conclusion}

In this paper we showed that there is a close connection between quantum 
Boolean operations and Reed-Muller expansions, which implies that a complete 
study on synthesis and optimization of quantum Boolean logic can be done 
within the domain of classical Reed-Muller logic. If we consider a positive polarity RM expansion 
and its corresponding quantum Boolean circuit, then using our proposed method we 
will get the same circuit efficiency we showed in \cite{younes1} {\it without the use of the 
truth table of the Boolean function or applying any transformations}.

In general the meaning of optimality is connected with practical constraints. For instance, the interaction 
between certain control qubits. Circuits depend on the physical implementation, it is sometimes difficult 
to take certain qubits as control qubits on the same $CNOT$ gates (involved in the same operation) because 
the interaction between these qubits may be difficult to control. Another constraint is the number of 
control qubits for a single $CNOT$ gate, at present it is not clear if the cost of implementation of multiple input 
$CNOT$ gates is higher than that of a fewer input $CNOT$ gates so it may be better to use fewer control 
qubits per $CNOT$ gate. Another constraint is the total number $CNOT$ gates in the circuit which should be 
kept to a minimum so they are able to maintain coherence during the operation of the circuit.

\end{document}